# Link Prediction and Navigability of Multiplex Energy Networks

Muhammad Kazim[1], Harun Pirim[2], Chau Le[3], Trung Le[4], and
Om Prakash Yadav [5]

[1]Ph.D student, Industrial & Manufacturing Engineering, North Dakota State University Dept. 2485, PO Box 6050, Fargo, ND 58102; e-mail: muhammad.kazim@ndsu.edu.
[2]Assistant Professor, Industrial & Manufacturing Engineering, North Dakota State University Dept. 2485, PO Box 6050, Fargo, ND 58102; e-mail: harun.pirim@ndsu.edu.
[3]Assistant Professor, Civil, Construction, and Environmental Engineering, North Dakota State University Dept. 2470, PO Box 6050, Fargo, ND 58102; e-mail: chau.le@ndsu.edu.
[4]Assistant Professor, Industrial & Management Systems Engineering, University of South Florida P.O Box 4202, FL 33620-000; e-mail: tqle@usf.edu
[5]Professor, Industrial & Systems Engineering, North Carolina Agriculture & Tech State University, P.O Box 408 NC 27411; e-mail: oyadav@ncat.edu

## ABSTRACT

In modern energy networks, where operational efficiency and resilience are critical, this study introduces an in-depth analysis from a multiplex network perspective—defined as a network where multiple types of connections exist between the same set of nodes. Utilizing Belgium's electricity and gas networks, we construct a five-layer multiplex network to simulate random node shutdown scenarios. We tailored the Jaccard and Adamic-Adar link prediction algorithms by integrating the concept of exclusive neighbors, thereby enhancing prediction accuracy with such multi-layered information. Emphasizing navigability, i.e., the network's ability to maintain resilience and efficiency under random failures, we analyze the impact of different random walk strategies and strategic link additions at various stages - individual layers, two-layer combinations, and three-layer combinations - on the network's navigability. Directed networks show modest improvements with new links, partly due to trapping effects—where a random walker can become circumscribed within certain network loops, limiting reachability across the network. In contrast, the undirected networks demonstrate notable increases in navigability with new link additions. Spectral gap analysis in directed networks indicates that new link additions can aid and impede navigability, depending on their configuration. This study deepens our understanding of multiplex energy network navigability and highlights the importance of strategic link additions, influenced by random walk strategies, in these networks.
**Keywords:** Multiplex Networks, Link Prediction, Jaccard, Adamic Adar, Random Walk, Energy Networks, Navigability, Network Analysis,

## INTRODUCTION

Modern economies increasingly rely on sophisticated energy networks, essential for operational efficiency, economic growth, and social welfare, thereby playing a pivotal role in these economies' robustness (Kabeyi and Olanrewaju 2022; Miśkiewicz 2020; Xu et al. 2021). Integrating sustainable energy sources into these networks is critical for aligning environmental objectives with economic feasibility and ensuring economic and infrastructural resilience (Albertus et al. 2020; Safari et al. 2019). Issues such as energy poverty and the imperative for efficient energy usage gain prominence, necessitating their thoughtful integration into network configurations and management strategies (Li et al. 2021). The addition of renewable energy sources, driven by environmental concerns, introduces new challenges, including the unpredictable nature of these sources and the need for network restructuring to support economic growth in various regions (Akrami et al. 2019; Xu et al. 2021).

While recent studies have begun to address the complex interdependencies within energy infrastructures, particularly from a multilayer network perspective(Pepiciello et al. 2023), no existing research has comprehensively explored the critical roles of link prediction and the strategic integration of new connections. These elements are essential for enhancing navigability—a measure of how efficiently energy flows through a network, which is crucial for maintaining network resilience and efficiency under stress (see 'Navigability Formulation' section). Our study bridges this gap by focusing on the strategic addition of links, which improves the network's ability to manage fluctuations and disruptions. By refining the Jaccard and Adamic-Adar link prediction algorithms to incorporate exclusive neighbors, we enhance prediction accuracy and, consequently, the network's structural and operational navigability across various layers (Bai et al. 2021). Detailed methodology and analysis of random walk strategies on navigability are explored in the subsequent sections.

## METHODOLOGY

Our study employs a comprehensive methodology to analyze a multiplex energy network modeled on Belgium's electricity and gas infrastructures. This network comprises 222 nodes, divided into electric nodes (IDs 01 to 75) and gas nodes (IDs 76 to 222). The primary focus of our dataset is on the source and target nodes and the energy flow between them. We have organized the network into five layers, each representing a unique shutdown scenario. These scenarios are generated using (SAInt) software simulating the knockout of random nodes within the electricity, the gas, or both networks. For a detailed description of the dataset and the preprocessing steps undertaken, please refer to the data documentation available on our GitHub repository, as linked in the Acknowledgment section. It is important to note that although our study constructs five scenarios, the number of potential scenarios in such analyses can vary significantly.

The methodology begins with trimming the dataset to refine the construction of the multiplex network. Subsequent stages include the identification of exclusive neighbors, covering the analysis of individual layers, two-layer combinations, and three-layer combinations. Following this, we apply customized link prediction algorithms, normalize the results, and assign weights to the predicted links. A key aspect of our analysis is examining the navigability or coverage of the original multiplex network before and after integrating these links from different stages. A detailed visual representation of this methodological approach is provided (see Figure 1).

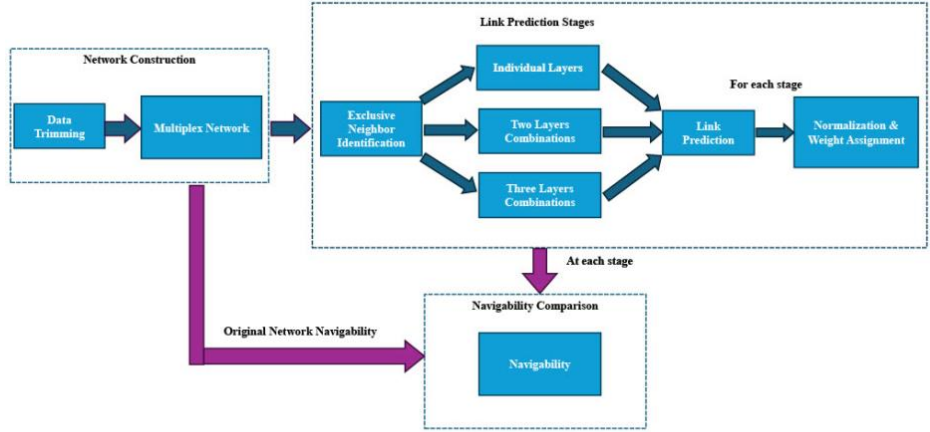

**Figure 1. Methodology flow chart**

**Data Trimming.** Edges with flow values below 90% of the maximum observed flow are removed to focus on the most significant network pathways and interactions and preserve data privacy.

**Multiplex Network Construction.** Following data trimming, we construct the multiplex network using a tensor-based framework (Aleta and Moreno 2019) . This involves defining the intralayer adjacency tensor $W_{\alpha\beta}^{(k)}$ for each layer *k*, representing connections within the same layer. The adjacency tensor is defined as $W_{\alpha\beta}^{(k)} = \sum_{i,j=1}^{N} w_{ij}^{(k)} E_{\alpha\beta}^{(ij)}$, where $w_{ij}^{(k)}$ denotes the connection strength between nodes *i* and *j* in layer *k*, and $E_{\alpha\beta}^{(ij)}$ is a basis component within the tensor space. This tensorial approach effectively captures the network's multidimensional complexity (see Figure 2 for our five-layer multiplex network illustration).

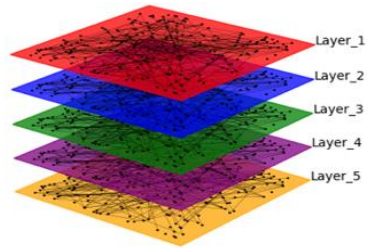

**Figure 2. Multiplex Energy Network**

Following the construction of the multiplex network, we compute its navigability. This computation is crucial for assessing the network's resilience—its ability to maintain functionality amidst potential disruptions. The navigability analysis sets the stage for the next critical step: identifying exclusive neighbors. This identification is a crucial preparatory phase for the forthcoming link prediction process, providing the necessary groundwork for our predictive modeling.

**NAVIGABILITY FORMULATION.** Navigability is critical in evaluating our multiplex network's functional integrity and robustness. We assess navigability using the coverage time equation ρ(t) to quantify network coverage over time (De Domenico et al. 2014). This incorporates the network initial condition, the transition matrix P, and the supracanonical vector $E_i^\dagger$. The coverage time is defined as:

$$\rho(t) = 1 - \frac{1}{N^2} \sum_{i,j=1}^{N} \delta_{i,j}(0) \exp[-P_j(0)PE_i^\dagger]$$

Here, $\delta_{i,j}(0)$ represents the network's initial condition (assuming the walker starts at vertex $j$). At larger time scales, influenced by the second smallest eigenvalue, $\lambda_2$, the coverage approximates to:

$$\rho(t) \approx 1 - \frac{1}{N^2} \sum_{i,j=1}^{N} \Delta_{i,j} e^{-C_{i,j}(1)t - C_{i,j}(2)\lambda_2^{-1}}$$

In this, $\Delta_{i,j}$ is the difference measure between nodes $i$ and $j$. with $C_{i,j}(1)$ and $C_{i,j}(2)$ being constants derived from the network's transition probabilities.

The network's dynamic changes are captured by the delta function, $\delta_{i,j}(t)$ evolving as:

$$\dot{\delta}_{i,j}(t) = -\delta_{i,j}(t)P(t)\mathcal{P}E_i^\dagger$$

Here, the probability vector $P(t)$ evolves according to the eigen decomposition of the normalized supra-Laplacian matrix $L$:

$$P(t) = P_j(0)e^{-Lt} = P_j(0) \sum_{l=1}^{NL} e^{-\lambda_l t} V_l$$

In this equation, L represents the normalized supra-Laplacian matrix, with $\lambda_l$ as its eigenvalues and $V_l$ as the corresponding eigenvectors.

Finally, the dynamical evolution of $\delta_{i,j}(t)$ is expressed as:

$$\dot{\delta}_{i,j}(t) = -\sum_{l=1}^{NL} e^{-\lambda_l t} C_{ij}(l)$$

We assess navigability for the original multiplex network and reevaluate it following the integration of predicted links at various stages, including single-layer, two-layer combinations, and three-layer combinations.

**Exclusive Neighbor Identification.** Exclusive neighbors are nodes connected to a target node within a specific subset of layers, emphasizing the importance of inter-layer connections (Berlingerio et al. 2013). The exclusive neighborhood of a node $v$ in layer subset D is defined as:

$$\text{NeighborsXOR}(v, D) = |\{u \in V \mid \exists d \in D: (u, v, d) \in E \land \nexists d' \notin D: (u, v, d') \in E\}|$$

This equation determines the count of neighboring nodes *u* that are exclusively linked to node *v* solely within layer subset D. By doing so; it highlights connectivity unique to specific layers.

**Link Prediction-Original Formulations and Enhanced Approaches.** Link prediction is a key aspect of our methodology. We begin with the foundational Jaccard and Adamic-Adar (AA) algorithms (Liben-Nowell and Kleinberg 2003). The Jaccard coefficient is calculated as the ratio of the intersection and union of the neighbors of two nodes. Mathematically, for nodes *u* and *v*, where $\Gamma(u)$ and $\Gamma(v)$ denote their respective sets of neighbors, it is expressed as:

$$Jaccard\ (u,v) = \frac{|\Gamma(u) \cap \Gamma(v)|}{|\Gamma(u) \cup \Gamma(v)|}.$$

On the other hand, the Adamic-Adar index assigns weights to shared neighbors based on the inverse logarithm of the degree of common neighbors. Its formula for nodes *u* and *v* is:

$$AA(u,v) = \sum_{w \in \Gamma(u) \cap \Gamma(v)} \frac{1}{log|\Gamma(w)|}$$

These original formulations act as a baseline for evaluating the likelihood of link formation within our multiplex network. We incorporate exclusive neighbors to adopt these algorithms for multiplex networks, leading to our modified versions.

**Modified Jaccard Algorithm for Multiplex Networks.** In this variant, we calculate the Jaccard score for each non-edge pair (*u*, *v*) by factoring in exclusive neighbours. This refines the link prediction process, better reflecting the unique interaction dynamics of each network layer.

**Algorithm 1:** Modified Jaccard Coefficient for Multiplex Networks

*Input: MultiplexGraph G*
*Output: Jaccard Scores for all non-edge pairs in G*
*1:* **procedure** *Modified Jaccard(G)*
*2:*    **Initialize** *J to be an empty list*
*3:*    **for each** *non-edge pair (u, v) in G* **do**
*4:*       **Let** $N_u$ *be the set of exclusive neighbours of u*
*5:*       **Let** $N_v$ *be the set of exclusive neighbours of v*
*6:*       **Let** *U be* $N_u \cup N_v$ *(Union of neighbours)*
*7:*       **Let** *I be* $N_u \cap N_v$ *(Intersection of neighbours)*
*8:*       **if** $|U| > 0$ **then**
*9:*          **Compute** *JaccardScore* $\frac{|I|}{|U|}$
*10:*         **Add** *(u, v, JaccardScore) to J*
*11:*   **end for**
*12:*   **return** *J*
*13:* **end procedure**

**Modified Adamic-Adar Algorithm for Multiplex Networks.** This algorithm is similarly adjusted to incorporate exclusive neighbors into the traditional Adamic-Adar calculation, summing the inverse logarithmic degree of each unique shared neighbor for the score. For a detailed description of the algorithm, see Appendix I. By emphasizing exclusive neighbors, our approach achieves a more nuanced understanding of link formation probabilities, enhancing the precision and relevance of link prediction in a multiplex environment.

**Normalization and Weight Assignment.** We normalize the prediction scores after applying the link prediction algorithms to ensure comparability and consistency across the network. Scores are scaled relative to the highest observed value, allowing us to uniformly evaluate link probabilities regardless of the inherent variance in original scores. We set a threshold of 0.5 for normalized scores, focusing our analysis on the most probable connections that exceed this threshold. During the weight assignment phase, these normalized scores are combined with average flow values between nodes, where the flow data is sourced from exclusive neighbor interactions. This integration results in weighted predicted links incorporated into the existing multiplex network. The weighting process prioritizes higher probability links and contextualizes them within the actual network dynamics, enhancing the realism and applicability of our simulation outcomes.

To evaluate the impact of these integrations, we recalibrate the network's navigability at different stages, including individual layers, two-layer combinations, and three-layer combinations. This recalibration process is crucial, as it allows for a systematic assessment of how new connections influence the network's overall performance and resilience, particularly regarding navigability.

**RESULTS AND DISCUSSION**

This section synthesizes our findings from applying the Adamic-Adar and Jaccard link prediction algorithms and navigability analysis within the multiplex energy network. We explore the impact of newly predicted connections on network navigability across various analysis stages, emphasizing their influence on enhancing navigability.

**A. Link Prediction Analysis:** Initially, the Adamic-Adar algorithm identified fourteen new connections, while the Jaccard algorithm found seven in individual layers. Progressing to two-layer combinations, these numbers increased significantly, with Adamic-Adar predicting 108 new links and Jaccard 76 demonstrating enhanced inter-layer connectivity. The complexity peaked in the three-layer combinations, with Adamic-Adar forecasting 387 and Jaccard 240 new links, respectively. Seventy unique links were ultimately identified throughout the various stages, demonstrating significant inter-layer connectivity, as depicted in Figure 3.

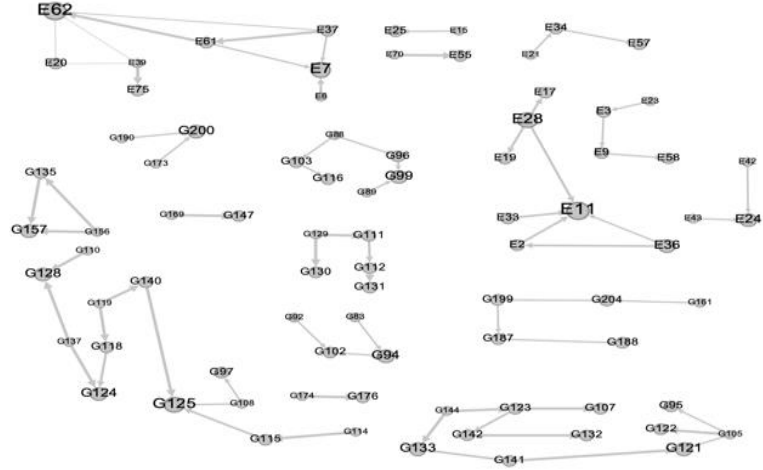

**Figure 3. Predicted Link Network:** Displays newly identified links between electric (E) and gas (G) nodes across various analysis stages. Nodes of larger size indicate a higher frequency of incoming links, suggesting their potential as pivotal hubs in the network.

The distribution of predicted links across the layers is detailed in Table 1A, Appendix II, highlighting the network's interconnectedness. This analysis enhances our understanding of the network's evolution and key connectivity areas. We next evaluate how these new connections affect network navigability, particularly focusing on traversal efficiency and coverage.

**B. Navigability Analysis**: This analysis evaluates how new link integrations across different stages and the network's directionality affect navigability using specific random walk strategies. We utilized the Classical Random Walk (RWC) strategy for directed and undirected networks. Detailed results for the Diffusive and PageRank strategies are documented in our GitHub repository.

Transition probabilities, crucial for these strategies, consider vertex transitions within and between layers, aiding in understanding network dynamics. These probabilities are detailed in Table 1, employing the notation from (De Domenico et al. 2014):

**Table 1: Transition Probabilities for Random Walk Processes in Multiplex Networks.**

| Transition | Classical Random Walk (RWC) | Diffusive Random Walk (RWD) |
|---|---|---|
| $P_{ii}^{\alpha\alpha}$ | $\dfrac{D_{(i)}^{\alpha\alpha}}{s_{i,\alpha} + S_{i,\alpha}}$ | $\dfrac{s_{\max} + D_{(i)}^{\alpha\alpha} - s_{i,\alpha} - S_{i,\alpha}}{s_{\max}}$ |
| $P_{ii}^{\alpha\beta}$ | $\dfrac{D_{(i)}^{\alpha\beta}}{s_{i,\alpha} + S_{i,\alpha}}$ | $\dfrac{D_{(i)}^{\alpha\beta}}{s_{\max}}$ |
| $P_{ij}^{\alpha\alpha}$ | $\dfrac{W_{ij}^{(\alpha)}}{s_{i,\alpha} + S_{i,\alpha}}$ | $\dfrac{W_{ij}^{(\alpha)}}{s_{\max}}$ |
| $P_{ij}^{\alpha\beta}$ | 0 | 0 |

These formulations consider transitions between vertices (denoted by Latin letters) and switching between layers (Greek letters). Here, *D* represents the diagonal matrix, and *W* is the adjacency matrix of a given layer. $s_{i,\alpha}$ is the intra-layer strength of vertex i in layer α. $S_{i,\alpha}$ is the total strength of vertex *i* in layer *α*. $s_{max}$ is the maximum strength across all vertices in the network.

**Original Multiplex**. In the examined multiplex network, the spectral gap, defined as the difference between the leading eigenvalue (λ1) and its subsequent eigenvalue (λ2) of the network's transition matrix, is a pivotal indicator of random walk efficiency. The undirected network variant exhibited a notably minimal spectral gap of $8.58 \times 10^{-5}$, culminating in an extended duration to achieve 90% coverage (26,823 units). This denotes a slower traversal rate, likely attributable to the inherent structural constraints within the network. Conversely, the directed network iteration, characterized by a more substantial spectral gap (0.00303), facilitated a swifter achievement of 90% coverage in merely 760.85 units. This more significant spectral gap signifies a markedly efficient random walk, indicative of rapid dispersion throughout the network, although potentially hindered by nodes lacking outgoing links.

**Stage 1-Individual Layers Links Integration.** After integrating new links from individual layers analysis, the undirected network's spectral gap decreased to $5.13 \times 10^{-5}$, and the 90% coverage time increased to 44,871. The directed network experienced a slight decrease in the spectral gap to 0.00297, with the 90% coverage time rising to 775.31.

**Stage 2-Two-Layers Combination Links Integration.** Adding these links reduced the undirected network's spectral gap to $1.17 \times 10^{-4}$, and coverage time extended to 1,959. The directed network spectral gap increased to 0.00307 with a decreased coverage time 749.

**Stage 3-Three-Layer Combination Links Integration.** With three-layer combination links, the undirected network's spectral gap decreases to $7.11 \times 10^{-4}$, with a 90% coverage time of 3,234. The directed network's spectral gap was reduced to 0.00267, and the 90% coverage time increased to 862. Figure 3 depicts the coverage in both directed and undirected networks, comparing the original network structure and subsequent additions of links using the classical random walk strategy.

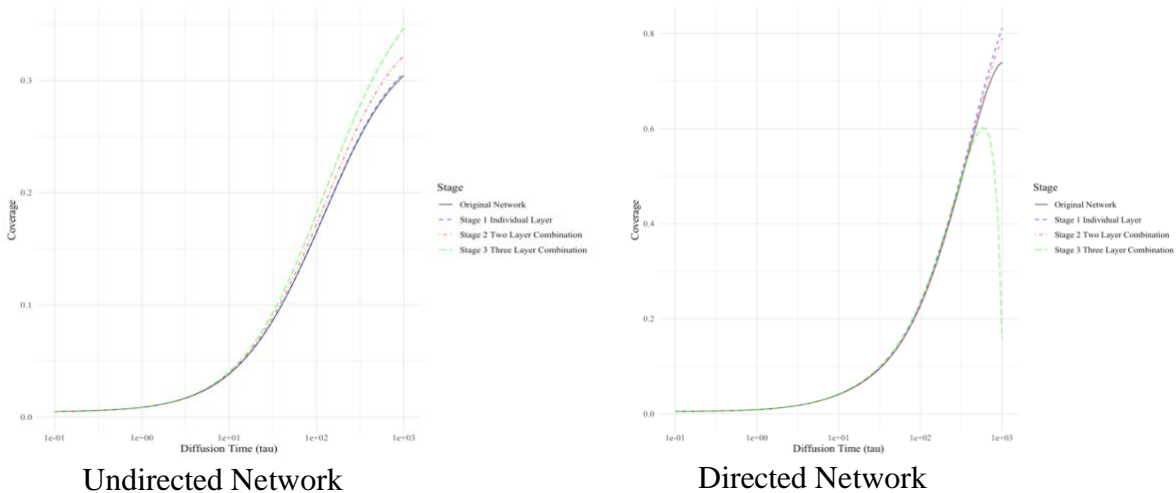

Undirected Network                    Directed Network

**Figure 4. Coverage Evolution in Directed and Undirected Networks using RWC.**

**C. Significance of Navigability Metrics**: The spectral gap directly correlates to network navigability. It quantifies the efficiency of traversal paths across the network, indicating how well it can sustain efficient operation amidst disruptions. Variations in this metric from staged link additions reveal insights into the balance between network robustness and traversal efficiency. This underscores the nuanced impacts of strategic link additions, particularly in improving or complicating navigability, especially in directed networks where new links might either resolve or introduce flow constraints.

By understanding these relationships, we better understand strategic link integration's role in optimizing network response and resilience. We employed three random walk strategies to analyze navigability: the Classical Random Walk (RWC) focuses on intra-layer transitions, the Diffusive Random Walk (RWD) examines the spread across multiple layers, and the PageRank strategy evaluates node centrality in a multi-layer context. Detailed results for the RWD and PageRank strategies and their specific contributions to our findings are provided in Appendix III.

**CONCLUSION**

Our study utilizing Belgium's electricity and gas networks demonstrates that strategic link additions significantly enhance multiplex network navigability. In undirected networks, we observed reduced times to achieve 90% coverage from 26,823 units initially to 3,234 units after implementing three-layer combination links, indicating a substantial increase in efficiency. While benefiting from new links, directed networks also showed varied responses due to their inherent structural constraints. These insights highlight the effectiveness of our tailored link prediction algorithms and random walk strategies in adapting to and improving network robustness under varied disruption scenarios. Such improvements are relevant to the specific network studied and suggest broader applicability for enhancing global energy systems. Future research should focus on diversifying network configurations and incorporating adaptive algorithms further to refine the resilience and efficiency of energy networks globally. These studies could extend our findings to dynamically manage network challenges in real-world settings.

**ACKNOWLEDGMENT:** The authors express their gratitude for the funding provided to support this study from the National Science Foundation (NSF) EPSCoR RII Track-2 Program under grant number OIA-2119691. The findings and opinions expressed in this article are those of the authors only and do not necessarily reflect the sponsors' views.

**GitHub**: https://github.com/CEL-lab/Link_Prediction_Multiplex/